\documentclass[amsmat,amssymb,amsfonts,aps,prl,twocolumn,showpacs]{revtex4}
\usepackage{graphicx}
\usepackage{dcolumn}
\usepackage{bm}
\newcommand{\beq}{\begin{equation}}  
\newcommand{\eeq}{\end{equation}}  
\newcommand{\beqa}{\begin{eqnarray}}  
\newcommand{\eeqa}{\end{eqnarray}}

\begin{document}
\title{Inelastic electron tunneling  spectroscopy of a single nuclear spin  }

\author{F. Delgado and J. Fern\'andez-Rossier}
\affiliation{Departamento de F\'{\i}sica Aplicada,
Universidad de Alicante, San Vicente del Raspeig, 03690 Spain }

\date{\today}

\begin{abstract}

Detection of a single nuclear spin constitutes an outstanding problem  in different fields of physics such as quantum computing or magnetic imaging. Here we show that the energy levels of a single nuclear spin can be measured by means of inelastic electron tunneling spectroscopy (IETS). 
 We consider two different systems, a magnetic adatom probed with STM and a single Bi dopant in a Silicon nanotransistor.     We find that the hyperfine coupling opens new transport channels which   can be resolved at experimentally accessible temperatures. Our simulations evince that IETS  yield information about the occupation of the nuclear spin states, paving the way towards transport-detected single nuclear spin resonance.

\end{abstract}

 \maketitle

Probing a single nuclear spin represents  both the ultimate resolution limit of magnetic resonance imaging and a requirement in quantum computing proposals where the nuclear spin is used as a qubit.  
The idea of storing and manipulating  information in nuclear spins goes back to quantum computing proposals based on P donors in Si\cite{Kane_nature_1998} and NMR quantum computing \cite{Gershenfeld_Chuang_science_1997}.  Because of their very small coupling to their environment,  the  nuclear spin coherence time is expected to be very long but, for the same reason,  quantum measurement of a single  nuclear spin remains a formidable task.  Recent experimental breakthroughs  have made it possible to perform single shot non-destructive measurement of a single nuclear spin by means of optically detected single spin magnetic resonance in NV centers in diamond \cite{Neumann_Philipp_science_2010}. 

Here we propose a setup based on  inelastic electron  tunneling spectroscopy (IETS) that would permit to probe the spin transitions of  a single nuclear spin with a space resolution down to  1 $\AA$   outperforming in this particular regard the optical detection.  Our proposal  builds   on  recent progress to probe  the spin of a single atom using two different strategies. On one side, Scanning Tunneling Miscrocope (STM)   inelastic electron  tunnel spectroscopy \cite{Heinrich_Gupta_science_2004,Hirjibehedin_Lutz_Science_2006,Hirjibehedin_Lin_Science_2007,Otte_Ternes_natphys_2008,Otte_Ternes_prl_2009,Loth_Bergmann_natphys_2010,Khajetoorians_Chilian_nature_2010} which allow to measure the electron spin spectral function of a single magnetic atom \cite{Rossier_prl_2009}  weakly coupled to a  conducting substrate.  On the other side,  the  fabrication of a Silicon nanotransitor where transport occurs through the electronic states of a single dopant\cite{Lansbergen_Tettamanzi_NanoLetters_2009,Tan_Chan_NanoLetters_2010}.

In STM-IETS experiment  electrons tunnel between the tip and the conducting substrate going through the magnetic atom. As the bias voltage $V$ is increased, a new conduction channel opens whenever $eV$ is larger than the energy of some internal excitation of the atom. In the case of isolated transition metal atoms with partially full $d$ shell, like Mn, Fe or Co,  the only internal excitations available in the range of a few meV are spin excitations associated to the magnetocrystalline anisotropy \cite{Hirjibehedin_Lin_Science_2007,Rossier_prl_2009}.   In the case of the single dopant nanotransistor,  the IETS of the electron spin in the donor level could be performed in the cotunneling regime\cite{Franceschi_Sasaki_prl_2001}. 

Exchange coupling to nearby magnetic atoms affects significantly the  spin excitation spectrum of the atom under the tip \cite{Hirjibehedin_Lutz_Science_2006,Rossier_prl_2009,Otte_Ternes_prl_2009,Loth_Bergmann_natphys_2010}. Hyperfine coupling to the nuclear spin should also result in a modification of the electronic spin spectral function which, in turn,  could be probed in IETS provided that the spectral resolution is high enough.
 Our mechanism differs from earlier theory work \cite{Berman_Brown_prl_2001,Balatsky_Fransson_PRB_2006} proposing to detect the nuclear spin  looking at its influence on the STM current noise spectrum.  Their approach is based on previous experiments where electronic spin fluctuations are detected in the   current noise spectra\cite{Manassen_Hamers_PRL_1989},  not in the conductivity spectra. 
Probing the spin transitions of a single nuclear spin would yield a completely unambiguous chemical identification of the atom and would be a first step towards transport-based quantum measurement of a single nuclear spin.

The rest of this letter is organized as follows.  We first describe the general theory that relates nuclear spin flips to IETS transport features.  Then we consider  the archetypical case of a single Mn atom in a Cu$_2$N surface \cite{Heinrich_Gupta_science_2004,Hirjibehedin_Lutz_Science_2006,Hirjibehedin_Lin_Science_2007,Loth_Bergmann_natphys_2010}. We find that
the detection of the nuclear spin excitations could be done at  $4$mK,  below  the recently demonstrated 10mK experimental 
limit~\cite{Song_Otte_nature_2010}, and how the visibility can be enhanced driving the nuclear spin out of equilibrium.
We then analyze the case of  $^{209}$Bi in  Silicon and we find it is an optimal system  to observe single nuclear spin flips at temperatures  up to 60mK. 

The electronic  spin $S$  and nuclear total angular momentum  $I$  are described by a Hamiltonian
${\cal H}_0(S,I)$ whose  eigenvalues and eigenvectors are denoted by $\epsilon_M$ and $|M\rangle= \sum_{I_z,S_z} \Psi_M(I_z,S_z) |I_z\rangle |S_z\rangle$, where $ |I_z\rangle |S_z\rangle$ is the basis in which both electronic and nuclear spin have well defined projection along the $z$ axis. The spin mixing coefficients $\Psi_M(I_z,S_z)$ depend on the specifics of the Hamiltonian, described below,  which includes hyperfine coupling,  Zeeman coupling and  magnetic anisotropy terms.  The inelastic transport spectroscopy is sensitive to transitions between states $M$ and $M'$  with excitation energy
 $\Delta_{M',M}=\epsilon_{M'}-\epsilon_M$. The electronic spin is coupled to electrons in the electrodes,  denoted as  tip (source) and the surface (drain) in the STM (nanotransistor) geometry, through  a
Kondo-like Hamiltonian\cite{Appelbaum_pr_1967,Rossier_prl_2009,Fransson_nanolett_2009,Delgado_Palacios_prl_2010}:
\begin{equation}
{\cal V}= \sum_{\alpha,\lambda,\lambda',\sigma,\sigma'} 
T_{\lambda,\lambda',\alpha}
\frac{\tau^{(\alpha)}_{\sigma\sigma'}}{2}  \hat{S}_{\alpha}
c^{\dagger}_{\lambda,\sigma} c_{\lambda'\sigma'}, 
\label{HTUN}
\end{equation}
where both electrode conserving and electrode non-conserving exchange couplings are included.
The operator  $c^{\dagger}_{\lambda,\sigma}$ creates an electron with spin $\sigma$ and orbital quantum number $\lambda=\eta,\vec{k}$,  where $\eta=T,S$ labels the electrode and $\vec{k}$ the wave vector. 
The index  $\alpha$ runs over $a=x,y,z$, with  $\tau^{(a)}$ and $\hat{S}_{a}$ the Pauli matrices
and  the  atom electronic  spin operators respectively. The $\alpha=0$ term (with $\tau^{(0)}$ and $\hat{S}_0$ the identity matrix) corresponds to potential scattering. 
We assume that exchange is  momentum independent, spin isotropic and electrode dependent:
  $T_{\lambda,\lambda',a} \equiv v_{\eta}v_{\eta'}{\cal T}$. Here  
  $v_{\eta}$ are dimensionless parameters that account for the  different coupling between the magnetic adatom and either the tip or the surface\cite{Delgado_Palacios_prl_2010,Delgado_Rossier_prb_2010}.

 The  inelastic current due to electronic-spin assisted tunneling $I_{\rm in}$  can be expressed\cite{Rossier_prl_2009} as a convolution of the electronic spin spectral function:
\begin{equation}
 {\cal S}(\omega)\equiv\sum_{M,M',a} P_M \left|\hat{S}_{a}^{M,M'}\right|^2
 \delta\left(\hbar\omega-\Delta_{M',M}\right)
 \end{equation}
 where  $P_M$ denotes the  average occupation of the state $M$
 and $\hat{S}_{a}^{M,M'}=\langle M|\hat{S}_a|M'\rangle$.
The inelastic current can be written as \cite{Rossier_prl_2009,Delgado_Rossier_prb_2010}:
\begin{eqnarray}
I_{in}(V)=\frac{g_{0}}{G_0} \sum_{M,M',a}
P_M(V)
 \left| \hat{ S}_{a}^{M,M'}\right|^2  i(\Delta_{M,M'}+eV).
\label{iinelast}
\end{eqnarray}
Here  $g_0\equiv\frac{\pi^2}{4}G_0 \rho_T \rho_S \left| {\cal T} v_T v_S\right|^2$ is the tunneling conductance, 
 $G_0=2e^2/h$ is the quantum  of conductance and $\rho_\eta$ the density of states of the $\eta$ electrode at the Fermi level. 
 The inelastic current associated to a single channel is given by $(e/G_0) i(\Delta+eV)= {\cal G}(\Delta+eV) - {\cal G}(\Delta-eV)$, with ${\cal G}(\omega)\equiv \omega\left(1-e^{-\beta \omega}\right)^{-1} $ and $\beta=1/k_B T$.

Importantly,  
  $\frac{di(\Delta+eV)}{dV}$ has a step at $eV=\Delta$ that accounts for the characteristic $dI/dV$ lineshape.
  Thus, whenever the bias energy $eV$ exceeds a transition energy between spin states with $\Delta_{M',M}=eV$ such that the initial state is occupied, $P_M>0$, and  the electronic spin flip transition is permitted, $\hat{S}_a^{M,M'}\neq0$, the differential conductance $dI/dV$ has a step and $d^2I/dV^2$ has a peak (or a valley at negative bias) with a thermal broadening\cite{Lambe_Jakelvic_PR_1968} of $\Gamma_{k_BT}\equiv 5.4 k_B T$.  Thus,  the $d^2I/V^2$  line-shape should be quite similar to the electronic spin spectral function $  {\cal S}(eV)$.  In the following we assume that the   intrinsic broadening of the spin excitations, due for instance to their coupling to the conducting substrate\cite{Delgado_Palacios_prl_2010,Gauyacq_Novaes_prb_2010,Delgado_Rossier_prb_2010},  is negligible compared to $\Gamma_{k_BT}$, as it happens in the case of Mn adatoms in Cu$_2$N \cite{Loth_Bergmann_natphys_2010,Gauyacq_Novaes_prb_2010,Delgado_Rossier_prb_2010}.

 The occupation functions $P_M$  can differ substantially from those of  equilibrium when the typical time elapsed between inelastic current events is shorter than the spin relaxation time \cite{Delgado_Palacios_prl_2010,Delgado_Rossier_prb_2010}.
We determine the occupation functions $P_M$  by solving a master equation\cite{Delgado_Palacios_prl_2010,Delgado_Rossier_prb_2010} that accounts for the dissipative dynamics of the current driven electronic spin interacting with the nuclear spin.  Both energy and spin are exchanged between the electrons in the electrodes  on one side, and the nuclear and electronic spin of the single atom on the other.  The scattering events involve the creation or annihilation of an electron-hole either in the same electrode, in which case no current is involved, or in different electrodes. Thus,  $P_M$ depends in general, on the voltage, the conductance and the temperature \cite{Delgado_Rossier_prb_2010}.

\begin{figure}[t]
\includegraphics[width=0.7\linewidth,angle=-90]{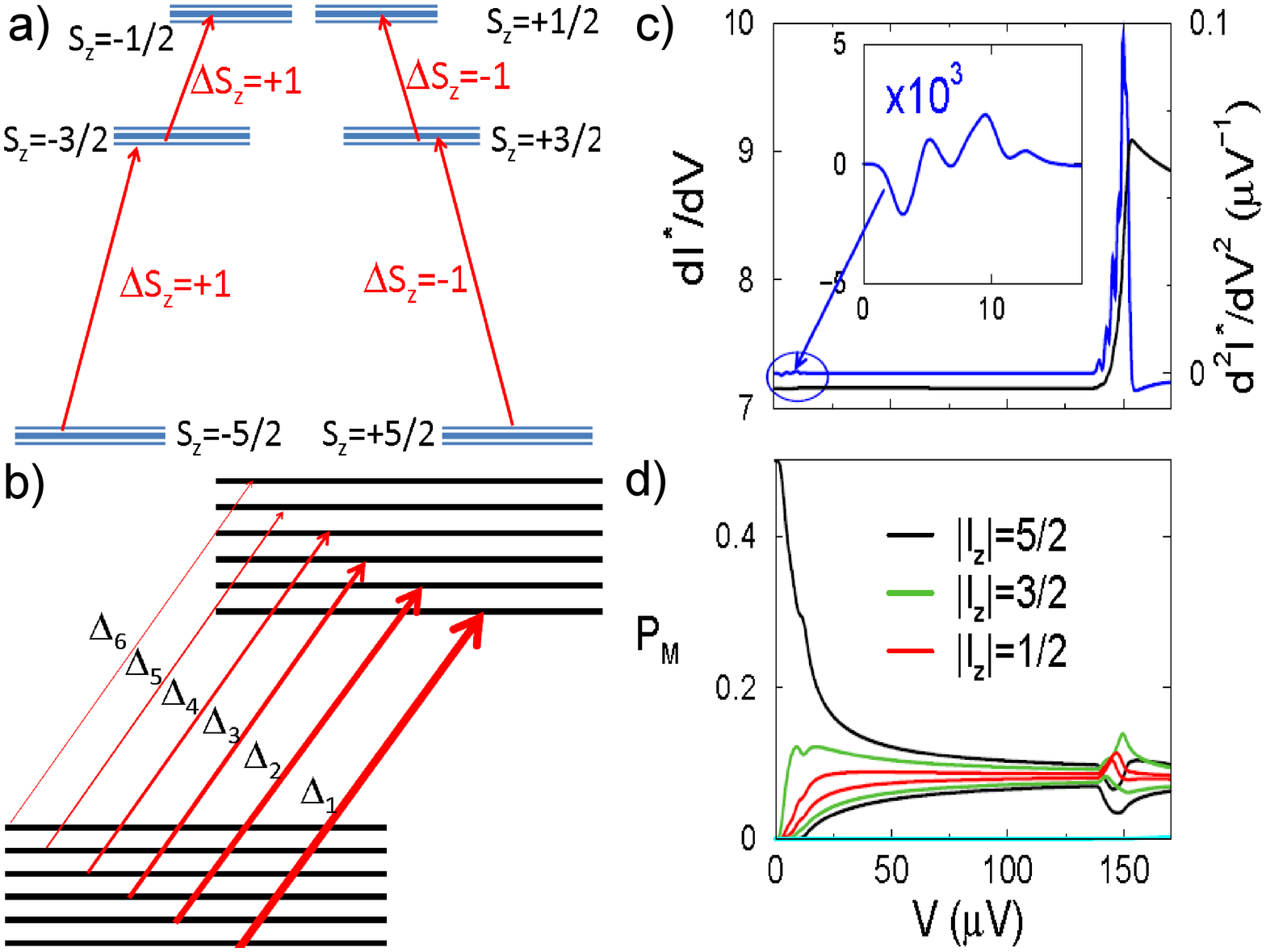}
\caption{ \label{fig1} (a) Energy level scheme of the $^{55}$Mn$^{2+}$ adatom on the Cu$_2$N surface. (b) Detail of the $S_z=5/2\to 3/2$ transition with resolved hyperfine structure.
(c) $dI^*/dV$ and  $d^2 I^*/dV^2$ in the high current regime $(v_T=v_S=1$) at $T=4$mK ($I^*=I/g_0$).  Inset: magnified low bias $d^2I^*/dV^2$. 
(d) Corresponding occupations of each of the 12 lowest energy electronuclear eigenstates of ${\cal H}_0({\rm S,I})$ for the $^{55}$Mn$^{2+}$ versus applied bias.
}
\end{figure}

 We now address the main question of this letter: under which conditions would IETS reveal transitions that provide information about the nuclear spin state?. We first consider the case of a single Mn adatom in a Cu$_2$N surface.  This system has been widely studied experimentally and theoretically\cite{Hirjibehedin_Lutz_Science_2006,Hirjibehedin_Lin_Science_2007,Fransson_nanolett_2009,Rossier_prl_2009,Persson_prl_2009,Loth_Bergmann_natphys_2010,Delgado_Palacios_prl_2010,Gauyacq_Novaes_prb_2010,Delgado_Rossier_prb_2010} but the role of the nuclear spin has been overlooked so far. The single atom electronic spin  can be described by means of a spin  $S=5/2$ Hamiltonian,
\beqa
\label{hchain}
{\cal H}_{\rm S}= D S_z^{2} + E(S_x^{2}-S_y^{2})+g_e\mu_B \vec{B}\cdot\vec{S}
\eeqa
where  $D=-39\mu$eV and $E=7\mu$eV  account for  the uniaxial and in-plane anisotropy respectively, whereas the third term describes the electronic Zeeman coupling.  At zero magnetic field the experimental\cite{Hirjibehedin_Lin_Science_2007}  $dI/dV$ features a step at an energy of $\Delta\simeq 4|D|$ associated to electronic spin flip between the two ground states which  neglecting $E$,  have  $S_z=\pm 5/2$ and the first excited states, with $S_z=\pm 3/2$, see Fig.~\ref{fig1}(a).  The only stable nuclear isotope of Mn is $^{55}$Mn   and has a nuclear spin of $I=5/2$, so that the electronic-nuclear system has 36 states in total.  The hyperfine structure associated to the coupling between $I$ and $S$ has been resolved with EPR and NMR experiments that address ensembles of more than 10$^{12}$ atoms, and can be described by a Heisenberg-type coupling. So, the  spin Hamiltonian reads as
\beqa
\label{hchain}
{\cal H}_0({\rm S,I}) = {\cal H}_{\rm S} +A\vec{S}.\vec{I}
\eeqa
The strength of the hyperfine coupling $A$ depends both on the nuclear magnetic moment and on the shape of the electronic cloud, which is environment dependent.  In the case of Mn, $A$ varies between 0.3 and 1$\mu$eV \cite{Walsh_Walter_PR_1965}.   Here we take $A=1\mu eV$. 
The effect of the hyperfine coupling is to split each of the 6 electronic levels into
 6 nuclear branches, as observed in Fig.~\ref{fig1}a,b).  The lowest electronic multiplet, corresponds to the $S_z=\pm5/2$,  
 while the $S_z=\pm 3/2$ and $S_z=\pm 1/2$ branches are found approximately at
 $4|D|$ and $6|D|$ above.
 
 The discussion can be simplified if we take advantage of the fact that $|D|\gg E,A$ 
  so that  the eigenstates  of ${\cal H}_0({\rm S,I})$ have, to zeroth order in $E$ and $A$, well defined projection of the electronic ($S_z$) and nuclear ($I_z$)  spin, although the numerical calculations are done with the exact states. The   energies  are approximately given by 
 $E_0(S_z,I_z)= D S_z^2 + A S_z I_z + g\mu_B S_z$.
  We find two kind of transitions:  low energy excitations with $\Delta S_z\simeq0$   and higher energy excitations with $\Delta I_z\simeq0,\Delta S_z=\pm 1$.The former are found at 
 small bias $eV\simeq S_z A \Delta I_z $, where $\Delta I_z$ is the change in nuclear spin. Whereas  these transitions  yield  very weak peaks in $d^2I/dV^2$ ,  shown in Fig.~1d), they contribute to drive the nuclear spin out of equilibrium, as seen in Fig. 3a).  

\begin{figure}[t]
\includegraphics[width=0.8\linewidth,angle=-90]{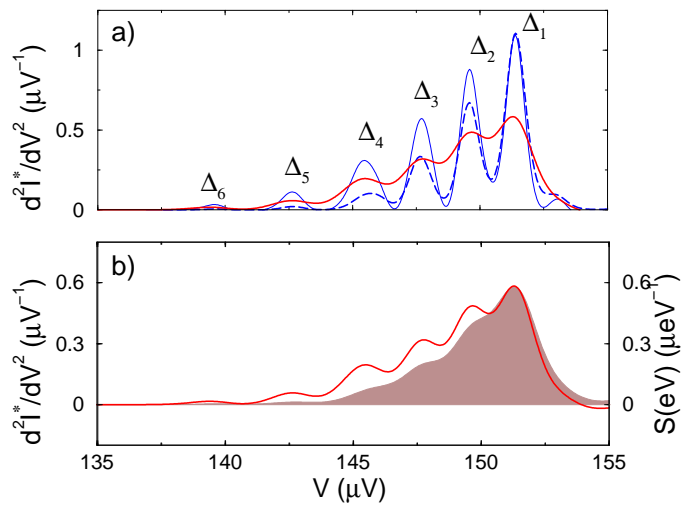}
\caption{ \label{fig2} (Color online) Conduction spectrum  $d^2I^*/dV^2$  as a function of applied bias ($I^*=I/g_0$).
(a) Spectra for $T=4$mK (thick solid line) and $T=2$mK (thin solid line) with $v_T=v_S=1$ (far from thermal equilibrium) and $T=2mK$  (thick-dashed line) with  $10v_T=v_S=1$ (close to thermal equilibrium). (b) $d^2I/dV^2$   for $T=4mK$ and $v_T=v_S=1$ (solid line) and spin spectral function $S(eV)$ as a function of applied bias (dashed line).}
\end{figure}

In contrast,  the  electronic spin transitions  between the electronic ground states $\pm 5/2$ and the first excited states $\pm 3/2$ that, to zeroth order, conserve the nuclear spin,  occur at higher $eV\simeq 4|D|$ and have a much stronger signal (Fig.  1c,d) and  2). At $B=0$ their energies are given by 
 $\Delta_{\pm}(I_z)= 4 |D|  \pm  A I_z $ 
so that, the hyperfine coupling  splits the lowest energy line of the electronic spin spectral function  into 6 lines separated by  $ A=\Delta_{\pm}(I_z)-\Delta_{\pm}(I_z\pm 1)$.  In order to have 
a thermal broadening smaller than $A$ ,  $k_BT$  must be reduced down to $\simeq 0.2 \mu$eV ($T\simeq 2 $mK), which would deplete
the thermal occupation of the higher energy nuclear states within the electronic ground state, and the visibility of the corresponding spin excitations. 
 In contrast, by decreasing the tip-atom distance \cite{Loth_Bergmann_natphys_2010} it is possible to drive the  system out of equilibrium  and  populate also the higher energy  nuclear spin states at low temperatures, as shown in Fig. 1d),  which makes it possible to observe all the transitions, as seen in Fig. \ref{fig2} even at $T=4mK$.

In figure  \ref{fig2}b) we plot  $d^2I/dV^2$ together with the the electronic spin spectral function ${\cal S}(eV)$, where the occupations $P_M$  evaluated at $eV$ and the delta function replaced by $\frac{d^2 i}{dV}$ . It is apparent that $d^2I/dV^2$ and ${\cal S}(eV)$  are  related.   Importantly, the spin spectral function contains information not only about the energy levels of the joint nuclear-electronic spin, but also about the occupations of the states. Thus,  the height of the $d^2I/dV^2$ can be correlated with the occupations of the nuclear spin states, outperforming current-noise spectroscopy\cite{Berman_Brown_prl_2001,Balatsky_Fransson_PRB_2006}. This indicates that it would be possible to use IETS-STM  as a detector in a  magnetic resonance experiment, in analogy with the optically detected single spin magnetic resonance. This would permit to probe the magnetic field with subatomic spatial resolution and accuracy afforded by the linewidth of  nuclear resonance experiment. 
 
\begin{figure}[t]
\includegraphics[angle=0,width=0.8\linewidth,angle=0]{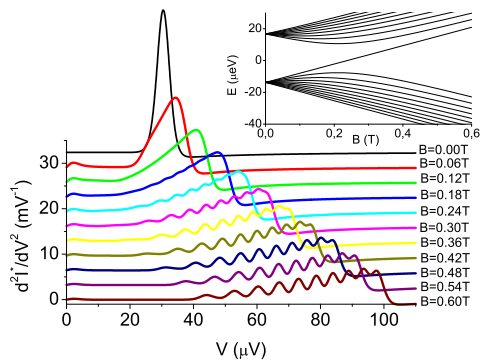}
\caption{ \label{fig3} (Color online)  Energy spectrum of $^{232}$Bi in Silicon as a function of applied field  and  the corresponding  $\frac{d^2I^*}{dV^2}$ spectra at $k_BT=10mK$.
}
\end{figure}
Nuclear spins  with larger  hyperfine coupling, like  Yb, Er or Pr,  
 \cite{Noel_Goldner_prb_2006,Ma_Yang_adndt_2004} could be probed with IETS at higher temperatures than Mn.  
 For instance  $^{167}$Er$^{3+}$ in Y$_2$SiO$_5$~\cite{Noel_Goldner_prb_2006} can have $A$  up to $6\mu$eV. 
The case of $^{232}$Bi in Silicon is particularly interesting.  
This system attracts  a lot of interest\cite{Morley_Warner_natmat_2010,George_Witzel_prl_2010} in the context of   quantum computing based on the nuclear spin of donors\cite{Kane_nature_1998} where  addressing the nuclear spin of a single dopant is required.  
The hyperfine coupling between the $I=9/2$ nuclear spin and the electronic spin of the donor state is quite large, $A=6.1\mu eV$ 
(1.48GHz)\cite{Morley_Warner_natmat_2010,George_Witzel_prl_2010}. 
 The zero field  Hamiltonian $ A\vec{I}\cdot\vec{S}$ can be diagonalized in terms of the total angular operator $F$, resulting in two multiplets (F=4,F=5)  with energy  $E(F)=  \frac{A}{2}\left(F(F+1)-S(S+1)-I(I+1)\right) $.
The zero field   splitting is as large as  $5A\simeq 30\mu $eV, which could be resolved  in IETS at $k_BT\simeq 60$mK.  The evolution of the spectrum
as a function of $B$ is shown in Fig.~3.   A single Bi dopant in Si could be probed 
by single spin IETS-STM , recently demonstrated in semiconductor substrates\cite{Khajetoorians_Chilian_nature_2010} or in a single dopant Silicon nanotransistor \cite{Lansbergen_Tettamanzi_NanoLetters_2009,Tan_Chan_NanoLetters_2010}, both in the sequential regime, for which the $dI/dV$ curve yield the single electron spectral function, with peaks at every one of the 20 energy levels \cite{Rossier_Aguado_prl_2007}, 
or in the cotunneling regime, described with the effective coupling (\ref{HTUN}) which yields information of the electronic spin spectral function.
 In figure 3 we show the
evolution of the  $d^2I/dV^2$ curves, as measured either with STM or in a nanotransistor in the  cotunneling regime,   for different values of $B$   at $k_BT=10$mK. It is apparent that transport  permits to  probe the spin-flip transitions between the Zeeman-split  states of  the joint electron-nuclear spin system.

In summary, we propose an experimental approach to probe a single nuclear spin using IETS of the hyperfine structure of the  electronic spin excitations, extending thereby the range of applicability of IETS. 
  Our simulations show that this technique yields information both about energy levels and the occupation of the nuclear spin states of a $^{55}$Mn adatom probed with a STM at $4$mK. 
  In the case of  a single $^{232}$Bi  dopant in a Si nanotransistor,  the hyperfine structure could be detected at
  $60$mK, well within range of current state of the art. 
  
   

We acknowledge fruitful discussions with C. Untiedt, C. F. Hirjibehedin and A. F. Otte.  This work was supported by MEC-Spain (MAT07-67845,  FIS2010-21883-C02-01, Grants  JCI-2008-01885 and  CONSOLIDER CSD2007-00010) and Generalitat Valenciana (ACOMP/2010/070).





\end{document}